\documentclass[english,a4paper,prx,twocolumn,amsmath,amssymb,superscriptaddress,longbibliography]{revtex4-1}
\usepackage{graphicx,color,soul}
\usepackage{mathtools}
\usepackage{enumerate}
\newcommand{\pd}[2]{\frac{\partial #1}{\partial #2}}
\newcommand{\mean}[1]{\langle #1\rangle}

\newcommand{\al}{\alpha}

\newcommand{\sig}{\sigma}
\newcommand{\Om}{\Omega}

\newcommand{\C}{\mathcal C}
\newcommand{\Ci}{\mathcal C^\text{is}}
\newcommand{\Nc}{\mathcal N}
\newcommand{\Tk}{T_\text{K}}     
\newcommand{\Ts}{T_\text{s}}     
\newcommand{\Tx}{T_\times}       
\newcommand{\Tc}{T_\text{c}}     
\newcommand{\tobs}{t_\text{obs}} 
\newcommand{\ta}{\tau_\alpha}    
\newcommand{\nr}{n_\text{rich}}  
\newcommand{\np}{n_\text{poor}}  
\newcommand{\phik}{\phi_\text{K}}    
\newcommand{\phir}{\phi_\text{rich}} 
\newcommand{\phip}{\phi_\text{poor}} 

\usepackage[dvipsnames]{xcolor}
\newcommand{\rev}[1]{{\color{black} #1}}

\newcommand{\ft}[1]{{\color{RoyalBlue} #1}}

\usepackage{array}
\newcolumntype{L}[1]{>{\raggedright\let\newline\\\arraybackslash\hspace{0pt}}m{#1}}
\newcolumntype{C}[1]{>{\centering\let\newline\\\arraybackslash\hspace{0pt}}m{#1}}
\newcolumntype{R}[1]{>{\raggedleft\let\newline\\\arraybackslash\hspace{0pt}}m{#1}}
\begin{document}

\title{Nonequilibrium Phase Transition in an Atomistic Glassformer: \\ 
the Connection to Thermodynamics}

\author{Francesco Turci}
\affiliation{H.H. Wills Physics Laboratory, University of Bristol, Bristol,
  BS8 1TL, UK}
\affiliation{Centre for Nanoscience and Quantum Information, Bristol BS8 1FD,
  UK}
\author{C. Patrick Royall}
\affiliation{H.H. Wills Physics Laboratory, University of Bristol, Bristol,
  BS8 1TL, UK}
\affiliation{Centre for Nanoscience and Quantum Information, Bristol BS8 1FD,
  UK}
\affiliation{School of Chemistry, University of Bristol, Bristol BS8 1TS, UK}
\affiliation{Department of Chemical Engineering, Kyoto University, Kyoto 615-8510, Japan}
\author{Thomas Speck}
\affiliation{Institut f\"ur Physik, Johannes Gutenberg-Universit\"at Mainz,
  Staudingerweg 7-9, 55128 Mainz, Germany}

\begin{abstract}
  \rev{Tackling the low-temperature fate of supercooled liquids is challenging due to the immense timescales involved, which prevent equilibration and lead to the {\emph{operational}} 
  glass transition. Relating glassy behaviour to an underlying, \textit{thermodynamic} phase transition is a long-standing open question in condensed matter physics. Like experiments, computer simulations are limited by the small time window over which a liquid can be equilibrated. Here we address the challenge of low temperature equilibration using trajectory sampling in a system undergoing a \emph{nonequilibrium} phase transition. This transition occurs in trajectory space between the normal supercooled liquid and  a glassy state rich in low-energy geometric motifs. Our results indicate that this transition might become accessible in \emph{equilibrium} configurational space at a temperature close to the so-called Kauzmann temperature, and provide a possible route to unify dynamical and thermodynamical theories of the glass transition.}
\end{abstract}

\maketitle


\section{Introduction}

Although statistical mechanics was firmly established more than a hundred
years ago~\cite{gibbs}, simple liquids remain a persistent challenge when
cooled to low temperatures. In particular, the dramatic super-Arrhenius
increase of the relaxation time so far eludes a generally accepted
explanation. A multitude of theoretical approaches has been
advanced~\cite{cava09,bert11
}, but obtaining data that enables
discrimination between these is challenging, due not least to the difficulties
in handling the huge timescales required to equilibrate 
supercooled
liquids. At some point (typically bypassing crystallization) the {structural} relaxation
time $\ta$ exceeds the experimentally or numerically accessible time scale and
the liquid falls out of equilibrium into a dynamically arrested state called a
glass. {This so-called \emph{operational 
 glass transition} is
  protocol-dependent and 
  {distinct from}
  equilibrium thermodynamic
  phase transitions}.

{However}, the idea that {at very low temperature a genuine
thermodynamic} phase transition {controls} 
dynamic arrest has been
around for a long time, starting with an observation by
Kauzmann~\cite{kauz48}: extrapolating the configurational entropy of the
liquid suggests that it should fall below that of the crystalline solid at a
finite (Kauzmann) temperature $\Tk$. One {resolution of} this apparent paradox
is to posit a thermodynamic transition to an ``ideal glass'' with very low
configurational entropy. {This transition is {hard to access} with the
{operational} 
glass transition intervening because of the accompanying
divergence of the structural relaxation time \cite{adam65}.}

More recent theories \rev{of the glass transition} continue to use the language of
phase transitions although they disagree on even the most basic
assumptions. On one hand the emergence of 
{slow dynamics}
in a rugged free-energy
landscape is anticipated through freezing, not to a single crystal but to a
vast number of random, aperiodic states~\cite{kirk89}; a picture that finds
justification from mean-field results obtained in higher
dimension~\cite{lubc07,char14} {and recent experiments \cite{albe16}}. 
{Approaches based on replica symmetry breaking also imagine a phase transition to a state where configurations are closely related to one another, {\emph{i.e.}} they have high overlap~\cite{pari10}.}
Another example, geometric frustration,
posits that {the population of} geometric motifs which minimise the local free energy (locally
favoured structures - LFS) 
{strongly increases}
upon supercooling and that the arrest seen is a manifestation of an avoided phase transition. Approaching the
ideal glass transition, the system may seek to minimise its entropy, which can
correspond to an increase in LFS~\cite{tarj05}.

Quite in contrast, in dynamical facilitation theory~\cite{chan10} dynamic 
correlations are the fundamental mechanism for the slow-down, with \rev{configuration-based} static
correlations being absent or at least irrelevant. {Originally developed with lattice models, evidence for the dynamical facilitation approach has now been obtained in experiment \cite{gokh14,chik14,pinc16}.} 
{Like the thermodynamic approaches noted above, }
the notion of a phase
transition is pivotal, but now the transition is between two dynamic regimes:
the supercooled liquid and a state with extremely slow
dynamics~\cite{hedg09}. The mathematical framework is that of statistical
mechanics combined with large deviations~\cite{touc09}. The transition shares
many similarities with the disorder-order transition of an Ising magnet; the
{crucial} conceptual difference being that configurations are replaced by
\emph{trajectories} (\textit{i.e.} time sequences of configurations) and the order
parameters {consider the time-integrated quantities such as} 
particle motion instead of magnetization.
{That is to say, the dynamical transition we consider (via a generalised or ``dynamical'' chemical potential) does not explicitly pertain to the motion of the particles, rather to a \emph{phase transition in trajectory space}, based on time-integrated quantities. What is meant by dynamical is that we consider time-integrated quantities over trajectories, but those quantities themselves may be based on static information, derived from configurations along the trajectory.} An intriguing prediction is the coexistence between the two dynamical regimes, which in lattice models is predicted to terminate in two critical
points {(one at high temperature and one at low temperature)}~\cite{elma10}. 
While numerical evidence supports coexistence and a first-order
nonequilibrium transition {in trajectory space} ~\cite{hedg09,spec12}, the putative lower critical
point remains out of reach for direct numerical investigations in atomistic
models. \rev{Whereas} the {operational} 
glass transition is a consequence of
the exceedingly long relaxation times of \rev{deeply supercooled liquids}, the
nonequilibrium phase transition in trajectory space is a genuine phase
transition, originating from non-analyticities in the derivatives of the
nonequilibrium equivalent of free energies (\textit{i.e.} the large deviation
functions of {time-integrated} observables~\cite{touc09}). {Recently, such dynamical phase transitions have become accessible to particle-resolved experiments \cite{chik14,pinc16}.}

{Returning to the idea of a} 
thermodynamic transition, {a key aspect} is some kind of structural change. While such structural changes appear to be minor when looking at two-point
measures like the structure factor, {higher-order measures reveal a richer behavior \cite{cosl07,lern09,mosa12,mali13fara,cubu15,pinn15}.}
{In particular, }the population of geometric motifs, so-called locally favoured structures \rev{(LFS)}, show a strong temperature-dependence~\cite{cosl07}. Indeed,
the non-equilbrium
transition to a state
with very low particle mobility can be
driven by increasing the LFS population~\cite{spec12b}, indicating that these
structural changes are not only a byproduct of cooling but play a crucial role
in the dynamic arrest. This finding is corroborated by simulations showing
that single particle motion and LFS are correlated~\cite{hock14}.

Numerical simulations are an indispensable tool to gain microscopic insights.
However, accessible time scales are still nine to ten decades away {in relaxation time} from the
{operational} 
glass transition temperature and more indirect methods have to be
devised to gain insight into the nature of the glass transition. For example, pinning (or confinement) of
particles~\cite{camm10,ozaw14,mart14} shifts the putative thermodynamic
transition into the time-window accessible to computer simulation, as has the
observation of distributions of overlaps in configurations of
particles~\cite{bert13a}. A further method to generate deeply supercooled,
equilibrated configurations is particle swaps~\cite{guti15,bert16}.

Here we introduce a new and powerful numerical method to
tackle the challenge of determining the low-temperature equilibrium behavior
of a popular atomistic glass former, {the Kob-Andersen binary
  mixture~\cite{kob94}. This model is loosely based on the metallic glass former
  Ni$_{80}$P$_{20}$.}  Unlike previous approaches, our
  method enables a connection between nonequilibrium phase transitions
  (\emph{i.e.} the approach of dynamic facilitation~\cite{chan10}) with more
  structure-based theoretical approaches which assume a thermodynamic phase
  transition. In particular, we use this method to extract configurations with
  exceptionally low configurational entropy and energy.

  Our numerical method exploits three effects: First, the central
  configurations of biased trajectories contain many locally favoured
  structures. Second, a higher population of LFS facilitates the sampling of
  configurations with low energies even at moderately supercooled
  temperatures. We develop methods to remove the simulation bias, \rev{which
    allow us to access not only} the simulated state point but \rev{also an
    extended temperature range} in its vicinity. The third effect is that the
  contribution to the total free energy from the local minima in the potential
  energy landscape and from the vibrational free energy
  decouple~\cite{scio99}. This allows us to access equilibrium properties at
  configurational temperatures $T$ (as opposed to the \rev{conventional thermodynamic}
  temperature) different {to} the sampling temperature $\Ts$.

Using the described method {(along with particle swaps \cite{grigera2001,bert16} adapted to the binary Kob-Andersen system with ``ghost insertion'' 
\cite{ashton2011})} 
we perform extensive biased simulations {of the} 
Kob-Andersen mixture, 
from which we construct part of the phase diagram for
the {\emph{time-integrated}} 
first-order transition from an LFS-poor phase (the normal liquid) to an LFS-rich phase. We find numerical evidence that coexistence between these phases is terminated at a finite temperature,
{implying} a lower critical point {eventually} accessible to the equilibrium system --- {were} it cooled sufficiently slowly. This scenario has significant
consequences because the dynamical transition at a higher temperature
{allows us to probe glassy configurations that are typical of much lower temperatures,}
while circumventing the prohibitive increase of equilibration times. We are thus able to probe the low-temperature fate of this 
model glassformer. 


\section{Methods}

\begin{figure*}
  \centering
  \includegraphics{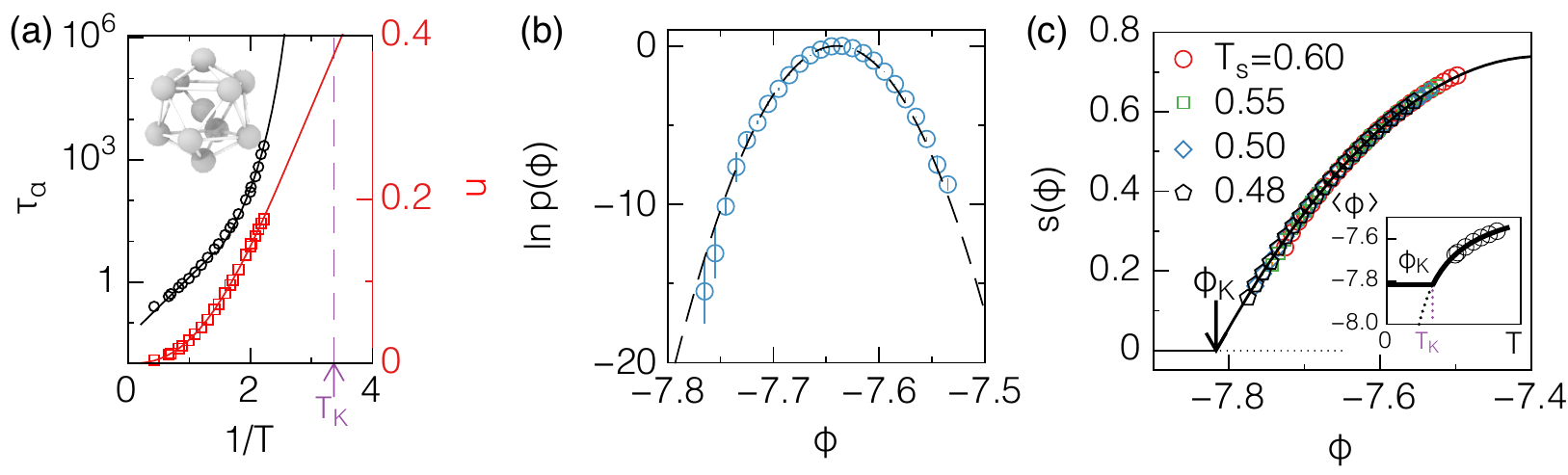}  
  \caption{Relaxation time and configurational entropy of the Kob-Andersen
    mixture. (a)~The relaxation time (black circles) strongly increases as a
    function of inverse temperature. {This is fitted with} the Vogel-Fulcher-Tamman
    expression $\ln\ta\propto(T-\Tk)^{-1}$ (solid line), which {we take to diverge} at
    $\Tk$ (arrow). Along with the relaxation time the average population
    $\mean{n}$ of LFS (red squares, see inset for a rendering of the
    structure) increases upon cooling. The red line is an empirical two
    parameter fit to Eq.~(\ref{eq:lfs}). (b)~The probability distribution of
    the inherent state energy $\phi$ is Gaussian (dashed line, shown for
    sampling temperature $\Ts=0.5$). (c)~Collapse of the configurational
    entropy $s(\phi)$ for three different sampling temperatures and fitted by
    a quadratic form (continuous line) down to very low energies. The energy
    $\phik\simeq-7.82$ at which the entropy would vanish is obtained from an
    extrapolation. In the inset, we use $\phik$ to estimate $\Tk\approx0.30$
    by extrapolating molecular dynamics results (black circles).}
  \label{fig:basics}
\end{figure*}

\subsection{Model}
\label{sec:model}

The Kob-Andersen binary mixture~\cite{kob94} consists of 80\% large particles
and 20\% small particles interacting through truncated and shifted
Lennard-Jones pair potentials. We employ the original potential
parameters. All numerical values are reported in Lennard-Jones units with
respect to the large particles, and we set Boltzmann's constant to
unity. Simulations are performed for a system of $N$ particles at number
density $N/V=1.2$ in a periodic box with constant volume $V$. We employ the
Andersen thermostat at sampling temperatures $\Ts=0.74,0.73,0.7,0.6,0.55,0.5,048$. Newton's
equations of motion are solved using Verlet's velocity algorithm with timestep
$0.005$~\cite{plimpton1995}. Trajectories are then stored as discrete
sequences $X=\{\C_{-K/2},\dots,\C_0,\dots,\C_{K/2}\}$ of configurations for
lengths $K=60$ and $K=100$. The time between successive configurations is
chosen such that trajectories have a physical duration of
$\tobs\approx 4.5\tau_{\alpha},7.5\tau_{\alpha}$, respectively.  Upon cooling
the structural relaxation time increases faster than exponentially
(super-Arrhenius behavior), see Fig.~\ref{fig:basics}(a). Whether the relaxation
time diverges at a finite temperature is still debated~\cite{heck08,elma10a}.

\subsection{Order parameters}

In the following we focus on two order parameters characterizing the
liquid. The first, \emph{{time-integrated}} order parameter
\begin{equation}
  \label{eq:N}
  \Nc[X] = \sum_{i=-K/2}^{K/2} n(\C_i)
\end{equation}
characterizes trajectories by counting the total number of particles in locally favoured structures,
where $n(\C)$ is the fraction of particles in LFS in configuration $\C$. For the specific
model considered here, the LFS corresponds to a bicapped square antiprism
formed by 11 particles as sketched in the inset of Fig.~\ref{fig:basics}(a). We
detect this motif employing the topological cluster classification
method~\cite{mali13fara,mali13tcc}. The temperature dependence of the LFS
population is fitted with the empirical Fermi function
\begin{equation}
  \label{eq:lfs}
  \mean{n} = \left[1+(T/T_{1/2})^\al\right]^{-1} \leqslant 1
\end{equation}
with fitted exponent $\al\simeq2.5$ and temperature $T_{1/2}\simeq0.25$ at
which the population extrapolates to $1/2$. The increase in LFS concentration
[cf. Fig.~\ref{fig:basics}(a)] has been shown to correlate well with the
propensity of particle mobility on lowering the temperature~\cite{hock14}.

Steepest descent quenches of the central configurations $\C_0$ to the local minima of the energy landscape yield the so-called inherent states~\cite{webe85}, which may be thought of as particle
configurations with the thermal noise removed. These inherent states $\Ci_0$
constitute the local minima in the energy landscape around which particles
vibrate before \rev{relaxing to} another local minimum. While in a perfect crystal
there would be only one inherent state, in the liquid there are many different
particle arrangements, and the number of these accessible amorphous inherent
states defines the configurational entropy. Our second, \emph{static} order
parameter is thus the inherent state energy (ISE) per particle
$\phi[X]=U(\Ci_0)/N$ of the \emph{central} configuration of trajectory $X$,
where $U(\C)$ is the potential energy of particle configuration $\C$ and $\Ci$
is the corresponding inherent state. {This is distinct in nature from the time-integrated order parameter in Eq. \ref{eq:N}.}

Practically, the ISE $\phi$ of central configurations $\C_0$ are obtained
using the FIRE algorithm~\cite{bitz06} limited to 1000 iterations in order to
make the generation of long sequences of trajectories computationally
feasible. Ground states of equivalent energies have been obtained employing
basin hopping techniques~\cite{wales1997} confirming that the reported values
for $\phi$ should be understood as upper (although tight) bounds to the true
inherent state energies.

\subsection{Biased simulations}

We go beyond the temperature regime that is accessible in conventional molecular dynamics
simulations. To this end biased simulations with $N=216$ and $N=400$ particles
are run at a moderately supercooled sampling temperature $\Ts$ to faster
explore phase space while at the same time sampling configurations with very
low potential energy and a high population of LFS {representative of low temperature}. To improve the sampling we
employ replica exchange~\cite{spec12}. We simultaneously extract {time-integrated} 
and static information {(central configuration)} by harvesting trajectories of length $\tobs$, which we
choose to be a few structural relaxation times $\tobs\approx 4.5\ta$ and
$\tobs\approx7.5\ta$ for runs with $K=60$ and $K=100$ configurations, respectively.

We employ the multistate Bennet acceptance ratio estimator~\cite{shir08,minh09} in order to calculate expectations and
distributions from the biased numerical data. For an arbitrary observable
$A[X]$, this amounts to evaluating the expression
\begin{equation}
  \label{eq:A}
  \mean{A}_{T,\mu} = \frac{\mean{A
      e^{-(1/T-1/\Ts)
      N\phi+\mu\Nc}}}{\mean{e^{-(1/T-1/\Ts)
      N\phi+\mu\Nc}}},
\end{equation}
where $\mean{\cdot}$ indicates the average over the sampled trajectories
  at sampling temperature $\Ts$. Here 
  $T$ {denotes} the
  target configurational temperature and $\mu$ the dynamical chemical
  potential.

In practice, our numerical approach allows us to improve the sampling of
trajectories {(and configurations)} that are rare for a given sampling temperature $\Ts$, explicitly
favoring trajectories with exceptionally large (or small) overall
concentrations of LFS $n$ through performing importance sampling in trajectory
space. Transition path sampling  \cite{dell02} is performed according to the structural bias in $n$ but any observable (including both $\phi$ and $n$) can be simultaneously tracked and its correct probability distribution, and in particular its mean value, can be recovered reassigning the correct weight to each trajectory, as
illustrated by Eq.~\ref{eq:A}.

\subsection{Temperature reweighting}
\label{sec:reweight}
The harvested trajectories also yield equilibrium expectation values beyond
the sampling temperature for observables $A[X]=A(\Ci)$ that depend on inherent
state configurations $\Ci$ only; provided that (i)~configurations are sampled
according to the Boltzmann weight $\propto e^{-U(\C)/\Ts}$ (at the sampling
temperature $\Ts$), and (ii)~vibrations are independent of the inherent state. Now, previous work \cite{scio99} has demonstrated via thermodynamic integration that for sufficiently low temperatures ($T<0.8$), the contribution of the inherent state energies to the free energy decouples from the vibrational contribution. {This satisfies condition (ii) while condition (i) is ensured by our sampling technique.}

{The key aspects of our technique are then as follows.} To show how it is possible to extract the equilibrium statistics of inherent state energies from the trajectory sampling {at temperatures {distinct to} $\Ts$, we split the total potential energy {into  two contributions:}
\begin{equation}
	U(\C)=N\phi(\Ci)+\delta U(\C|\Ci),
\end{equation}
{where $\delta U(\C|\Ci)$ indicates the extra potential energy of a configuration $\C$ compatible with the inherent state configuration $\Ci$.}

{Using such a decomposition, we can evaluate the thermal average $\langle\cdot\rangle$ of the following expression:}
\begin{align}
\label{eq:rew1}
	  \mean{A e^{-\vartheta N\phi}}  & = \sum_{\Ci}\sum_{\C|\Ci} A(\Ci)e^{-\vartheta N\phi}
  \frac{e^{-U/\Ts}}{Z(\Ts)}\\
	  &=\sum_{\Ci}\sum_{\C|\Ci} A(\Ci)e^{-(\frac{1}{T}-\frac{1}{\Ts}) N\phi}\frac{1}{Z(\Ts)}e^{-\frac{N\phi+\delta U}{\Ts}}
\end{align}
{with partition sum $Z(T)=\sum_\C e^{-U(\C)/T}$ and $\vartheta=1/T-1/\Ts$. We then define the \textit{restricted}} partition sum 
\begin{equation}
	\tilde Z(T|\Ci)=\sum_{\C|\Ci}e^{-\delta U/T},
\end{equation}
sampling the fluctuations out of the inherent state $\Ci$. If
the restricted partition sum is approximately independent of the inherent
state, $\tilde Z(T|\Ci)\approx\tilde Z(T)$, we {can rewrite Eq.~\eqref{eq:rew1} as}
\begin{equation}
	 \mean{A e^{-\vartheta N\phi}} = \frac{\tilde Z(\Ts)}{Z(\Ts)}\sum_{\Ci}A(\Ci)e^{-N\phi(\Ci)/T}.
\end{equation}
{Finally, this expression can be used to compute the reweighted average $\langle\cdot\rangle_{T,\mu}$ evaluated at equilibrium $\mu=0$ in Eq.~\eqref{eq:A} and obtain}
\begin{equation}
  \mean{A}_{T,\mu=0} = \frac{\sum_{\Ci}Ae^{-N\phi/T}}{\sum_{\Ci}e^{-N\phi/T}}
\end{equation}
corresponding to the equilibrium expectation of $A$ at temperatures $T\neq\Ts$
different from the sampling temperature.

{\subsection{Ghost particle Monte-Carlo}
In order to compare our trajectory sampling results with an alternative equilibrium technique, we perform Monte-Carlo simulations where we perturb the Hamiltonian  of the system in order to favor local relaxation of caged particle arrangements. 

The method is inspired {by} 
ghost particle insertion \cite{ashton2011}  and swap Monte-Carlo algorithms \cite{grigera2001,bert16}. In our implementation, a subset of the particles in the system is associated {with} a stochastic variable that determines the maximum repulsive forces acting on the particles of the subset. This facilitates local cage-breaks and relaxation. The particles {which interact via this} capped potential are reselected randomly after 
{a certain number of }
iterations are called ``ghost'' particles, hence the name of the method. 
The random walk of the stochastic variable satisfies detailed balance and includes states for which the Hamiltonian is {that} 
of the unperturbed system. We record those states and compute the average inherent state energy $\langle \phi\rangle$ from an equilibrium distribution of configurations \cite{ashton2011}. For more details on the method and the algorithm, see Appendix~\ref{app:gMC}.}

\begin{figure}[t]
  \centering
  \includegraphics{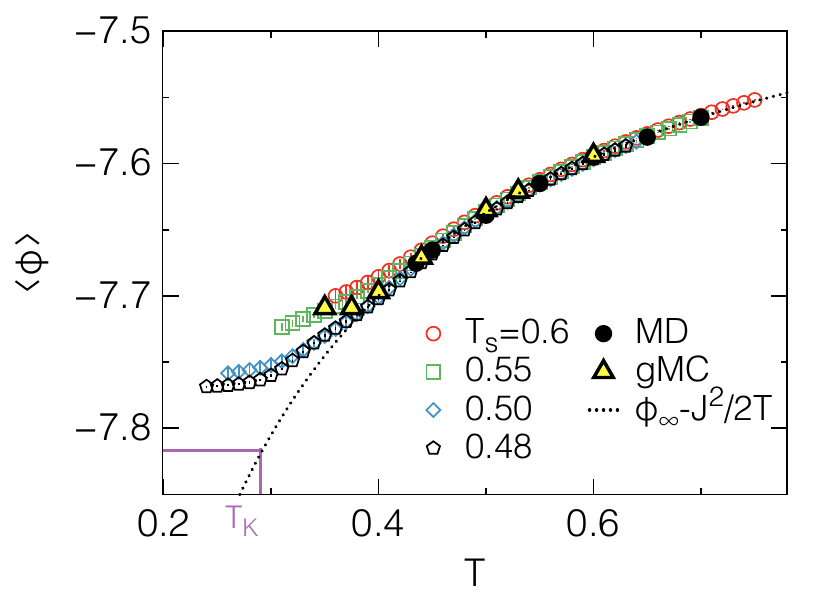}
  \caption{Importance sampling of inherent state energies at sampling
    temperatures $\Ts=0.60,0.55,0.50, 0.48$. The dotted line indicates the
    prediction of Eq.~\ref{eq:mean} for the equilibrium liquid, while filled
    {circles} represent the values obtained from unbiased Molecular Dynamics
    simulations {and yellow triangles indicate the equilibrium values obtained via a ghost {particle} Monte Carlo (gMC) technique (see
    {text and} Appendix~\ref{app:gMC})}. Deviations from the equilibrium liquid occur only at low $T$
    and become smaller if the sampling is performed at lower $\Ts$. Also
    indicated is the putative value of the Kauzmann temperature $\Tk$ and the
    corresponding energy of the equilibrium liquid. {Notice that trajectory sampling at $T_s=0.50, 0.48$ outperforms the longest Monte Carlo calculations at the lowest temperatures, reaching much deeper inherent state energies. {In particular the lowest temperature estimate of the inherent state energy in the gMC shows that also the Monte Carlo technique can remain trapped in long-lived metastable states similar to the ones obtained via trajectory sampling at high $\Ts$.}}}
  \label{fig:phi}
\end{figure}

\section{Results}

{Our results are twofold. First, we describe our reweighting procedure to obtain configurations representative of exceedingly low temperature, far beyond the regime accessible to conventional techniques. This shows that the configurational entropy and inherent state energy are both compatible with a transition to a state of very low entropy at a finite temperature $\Tk$. We then proceed to consider the behavior of the system in the space of the nonequilibrium phase transition $(T,\mu)$. We show that the nonequilibrium phase transition appears (upon extrapolation) to be bound from below, with a lower critical point $\Tc$. This lower critical point lies close to $\Tk$ (or perhaps even at $\Tk$). We discuss how this proximity of the lower critical point and $\Tk$ may allow some union of the different approaches to the glass transition.}

\subsection{Configurational entropy}

We first focus on the behavior of the central configurations. Our simulations
sample configurations with a wide range of \rev{inherent state energies}, which we compile into
distributions $p(\phi)$ for the different sampling temperatures. In agreement
with previous studies~\cite{buch99,scio99}, we find that these distributions
are well described by a Gaussian, see Fig.~\ref{fig:basics}(b). Moreover, it has
been demonstrated that at low enough temperature (including the sampling
temperature $0.48\leqslant\Ts\leqslant0.7$ used here) the vibrational free
energy is independent of $\phi$ to a very good approximation~\cite{scio99}.

Let $\Om(\phi)\propto\Om_\infty e^{N\sig(\phi)}\delta\phi$ be the number of
amorphous inherent states with energy per particle $\phi$ within an
interval $\phi\pm\delta\phi/2$. Here, $\sig(\phi)$ is the enumeration function
and $\Om_\infty\simeq e^{Ns_\infty}$ is the maximal available volume in
configuration space {\emph{i.e.} the high temperature case where $s_\infty$ is the configurational entropy.}
The (configurational) temperature $T$ corresponds to the
inverse slope, $d\sig/d\phi=1/T$. In the limit of large $N$, the number $\Om$
is either extensive or becomes exponentially small. Hence, in the
thermodynamic limit, the extensive configurational entropy becomes
$\ln\Om(\phi)\simeq Ns(\phi)$ with $s(\phi)=s_\infty+\sig(\phi)$ for
$\phi\geqslant\phik$ and $s(\phi)=0$ for extreme {inherent state} energies
$\phi<\phik$. {Here $\phik\simeq-7.82$ is the inherent state energy of the system at \rev{the Kauzmann temperature}
  $\Tk$.}

{The Gaussian shape of $\Om(\phi)$ implies that the 
enumeration function we extract 
$\sig(\phi)$ should be quadratic,}
$\sig(\phi)=-(\phi-\phi_\infty)^2/J^2$ from the measured distribution with
fitted maximal energy $\phi_\infty\simeq-7.38$ and \rev{energy scale} 
$J\simeq0.502$. The
resulting configurational entropy $s(\phi)=s_\infty+\sig(\phi)$ is plotted in
Fig.~\ref{fig:basics}(c) using the value $s_\infty\simeq0.74$ reported
previously~\cite{scio99}. Our numerical scheme is thus able to cover a wide
range of inherent states and the excellent agreement demonstrates that the
configurational entropy of the liquid is indeed very well described by a
quadratic function.

For the thermal average of the inherent state energy one finds
\begin{equation}
  \label{eq:mean}
  \mean{\phi} = \phi_\infty - \frac{J^2}{2T}
\end{equation}
as a function of temperature $T$. 
\rev{While we make the distinction between
  configurational and conventional thermodynamic temperature, for quantities based on
  inherent states, such as the inherent state energy itself, the
  configurational temperature corresponds to the conventional
  temperature.}


The scenario that $\mean{\phi}=\phik$ is reached at a finite temperature $\Tk$ is shown in Fig.~\ref{fig:basics}(c) by extrapolating simulation data. Figuratively speaking, at this temperature the liquid would ``run out'' of amorphous configurations and would undergo a thermodynamic phase transition to an ideal glass with constant inherent state energy $\phik$ and very low entropy. Although the existence of such an ideal glass state is debated on several grounds, e.g.~\cite{stil88,done07}, we keep $\phik$ as a convenient estimate for the lowest energy accessible to amorphous configurations.  Extrapolating the quadratic form for $s(\phi)$ to lower energies, one derives $\Tk=J/(2\sqrt{s_\infty})$ for the Kauzmann temperature yielding $\Tk\simeq0.3$ for the present system in agreement with previous estimates {ranging from $0.297<\Tk<0.325$}~\cite{scio99,colu2000,spec12b}. This value is compatible with an extrapolation of molecular dynamics data for the average ISE $\mean{\phi}$, which we show in the inset of Fig.~\ref{fig:basics}(c). Comparing with the increase of $\ta$ [Fig.~\ref{fig:basics}(a)], it is clear that $\Tk$ is far below temperatures for which the liquid can be equilibrated in computer simulations.

In Fig.~\ref{fig:phi} the average {inherent state energy} 
$\mean{\phi}_{T,\mu=0}$ is shown as a
function of (configurational) temperature $T$ employing 
{our} scheme,
showing good agreement with data obtained from molecular dynamics
simulations {and with data obtained from the 
ghost {particle} Monte Carlo technique that promotes relaxation via the usage of staged, capped potentials (see Appendix \ref{app:gMC}). We observe that the results from trajectory sampling for the lowest sampling temperatures are in good agreement with Eq.~\eqref{eq:mean} down to $T=0.37$, where the relative deviations start to exceed 
$0.1\%$. At $T_{\rm s}=0.48, 0.50$, the average inherent state energy departs from Eq.~\eqref{eq:mean} for $T<0.37$ and glassy, amorphous, very low energy states are obtained. These are inaccessible to conventional techniques since the relaxation time is expected to be, via extrapolation, $\tau_\alpha(T=0.37)>10^4\tau_{\alpha}(T_{\rm MC}=0.435)$}. Assuming that the Kauzmann energy $\phik$ represents the
``bottom'' of the energy landscape of amorphous states, then we expect that
for $T<\Tk$ there should be no further reduction in $\mean{\phi}$, which is
indeed supported by our data for successively lower sampling temperatures
$\Ts$.


\subsection{
{Trajectory space} phase transition from LFS-poor to LFS-rich}

\begin{figure*}[t]
  \centering
  \includegraphics{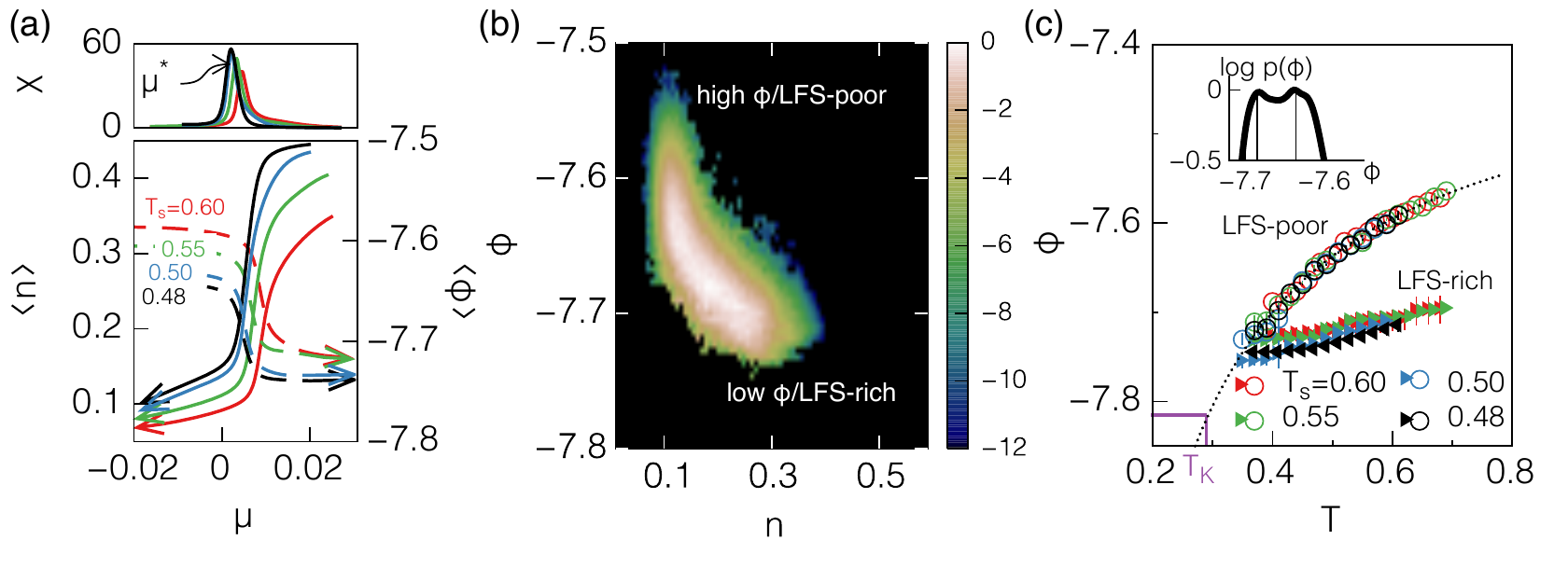}
  \caption{The inherent state energy drops across the dynamical
    transition. (a)~Signature of the dynamical phase transition: the average
    trajectory population $\mean{n}_{\Ts,\mu}$ of LFS (continuous lines) shows
    a sudden increase at a non-zero $\mu$. Top, the susceptibility $\chi(\mu)$
    quantifying the magnitude of fluctuations {in LFS populations} is shown, the peak of which
    defines $\mu_\ast$. Around the same $\mu_\ast$, the average ISE
    $\mean{\phi}_{\Ts,\mu}$ of central configurations (dashed lines) drops
    substantially. (b)~ Logarithm of the joint probability of LFS population $n$ and ISE $\phi$,
    for $\Ts=0.55$ and $\mu=0.005\approx0.7\mu_\ast$. (c)~Typical energies of
    the two populations as a function of temperature: LFS-poor states
    (circles) essentially follow the equilibrium {liquid} 
    curve (dotted line)
    while LFS-rich ones (filled triangles) are characterised by much lower
    energies, with a $\Ts$-dependent tail at low temperatures. Inset: ISE
    probability distribution $p(\phi)$ evaluated at $\mu=\mu_\ast$ for
    $T=0.55$.}
  \label{fig:dist}
\end{figure*}

We now consider the full trajectories {rather than the central configuration as above}. Formally, one can treat the particles
in LFS and {non-LFS (free)} particles as two chemical species with the same chemical
potential that interconvert freely. The extension to ensembles of trajectories
is, at least formally, straightforward with $\mu$ the dynamical analogon of
the chemical potential difference. Unbiased equilibrium dynamics 
corresponds to $\mu=0$. In line with physical intuition, positive $\mu>0$
gives trajectories with larger population a higher weight, which is
demonstrated in Fig.~\ref{fig:dist}(a). Here we show the average population
$\mean{n}_{\Ts,\mu}$ of LFS as a function of $\mu$ for the three sampling
temperatures. At some $\mu_\ast>0$ there is a sharp increase of the population
indicating a transition from the normal supercooled liquid (LFS-poor) to a
state composed of trajectories 
with 
a large number of LFS
(LFS-rich). Indeed, it has been shown that the transition becomes sharper for
larger systems~\cite{spec12b}, where ``larger'' can be both larger $\tobs$
(longer trajectories) and larger $N$ (more particles while holding the density
constant). This implies a first-order transition in trajectory space, where the jump of the order
parameter (the population of LFS) is rounded by finite-size
effects~\cite{challa1986}. Practically, the value of {$\mu$ at which the transition occurs,}  $\mu_\ast$ is determined
from the peak of the susceptibility
\begin{equation}
  \label{eq:chi}
  \chi(\mu) = \pd{\mean{n}_{\Ts,\mu}}{\mu}
\end{equation}
maximizing the fluctuations of the order parameter $n$ {as shown in Fig. \ref{fig:dist}(a)}. The susceptibility
measures the sensitivity of the population {of LFS along the trajectory} {to a change in the field strength $\mu$. This} 
becomes
very large close to the phase transition and is expected to diverge in the
limit of infinite system size. Note that decreasing the sampling temperature,
the peak position moves towards $\mu=0$, which corresponds to the unbiased,
equilibrium 
system, {\emph{i.e.} that encountered in experiment}. \rev{Moreover, {upon decreasing temperature}, the peak height
$\chi_\ast=\chi(\mu_\ast)$ grows. We return to the intriguing question of the low
temperature behavior of $\mu_\ast$ below. }

Having identified an LFS-poor and an LFS-rich phase, in Fig.~\ref{fig:dist}(a)
we show that the transition is accompanied by a complementary drop of the inherent state energies
of central configurations~\cite{jack11}. Indeed, looking at the joint
distribution of $n$ and $\phi$ in Fig.~\ref{fig:dist}(b), two distinct
populations of trajectories can be {identified.} 
In particular, trajectories with
a large number of LFS {feature }
central configurations that typically have low
inherent state energy. 

From the marginal distribution of ISE [inset of
Fig.~\ref{fig:dist}(c)] we extract the two typical values $\phip$ and $\phir$
for LFS-poor and LFS-rich states, respectively, as a function of
configurational temperature $T$, using Eq.~\ref{eq:A}. These we plot in
Fig. \ref{fig:dist}(c). The inherent state energies for the LFS-poor liquid follow the
prediction of Eq.~(\ref{eq:mean}) for the average {inherent state energy} \rev{and are therefore indistinguishable, up to our numerical precision, from the equilibrium liquid}. 
In contrast, the much
lower $\phir$ shows a more complex behaviour: at high temperatures, the
different $\Ts$ converge at the same energy, which is a slow monotonically
increasing function of the configurational temperature $T$. For the lowest
temperatures $T$, there is a systematic dependence on the sampling temperature
$\Ts$, with lower $\Ts$ allowing to better sample lower energy states and thus
leading to lower average ISE. Due to the much weaker dependence on $T$ of
$\phir$, the energy gap $\phip-\phir$ is reduced with decreasing $T$, so that
it suggests the existence of a finite temperature at which the
two energies reach the same value and the gap vanishes. \rev{In other words,
  we expect a low but finite temperature at which the equilibrium liquid becomes indistinguishable from the low-energy, LFS-rich phase.} 

\subsection{Nonequilibrium phase diagram}

\begin{figure*}[t]
  \centering
  \includegraphics{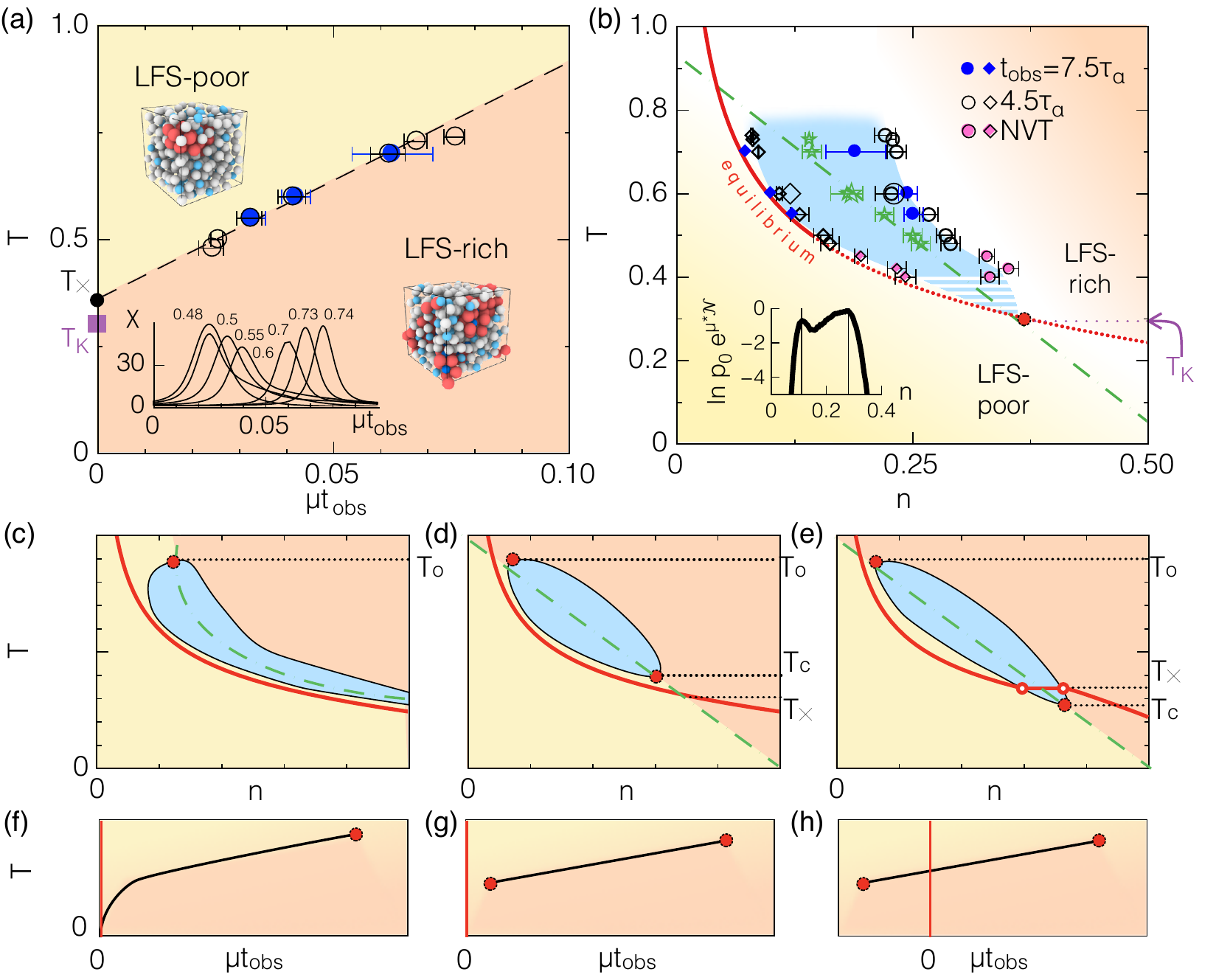}
  \caption{Phase diagram {for the dynamical transition} and possible scenarios for a lower critical point. (a) Coexistence in the $T-\mu$ phase
    diagram, with $\mu_\ast$ scaled by the trajectory length. Linear
      extrapolation implies { $\Tx(\mu_\ast\to0)=0.37$ (black dot) close to $\Tk=0.30$} (purple
    square). {Inset: susceptibilities $\chi$ of the LFS population labelled by their sampling temperatures for a system for $N=216$ and $\tobs=4.5\tau_\alpha$, showing a} {non-monotonic behaviour with increasing peak heights at low and high temperatures}. Two snapshots of LFS-rich and LFS-poor configurations are also illustrated,
    with the non-LFS particles in white and pale blue. (b) Coexisting populations for $N=216$ particles and several
    sampling temperatures (filled symbols $\tobs\approx7.5\ta$, empty symbols
    $\tobs\approx4.5\ta$) and $N=400$ at $\Ts=0.6$ (larger empty
    symbols). Green stars denote $n_\ast(T)=\mean{n}_{T,\mu_\ast}$ fitted to a
    linear function (dashed-dotted green line). {Additionally, coexisting populations at temperatures lower than $T=0.48$  are obtained from nonequilibrium molecular dynamics at constant temperature (NVT) with LFS-rich initial conditions, as discussed in the Appendix.} A
    possible coexistence region is sketched (shaded area). The unbiased
    equilibrium dynamics (red continuous line, Eq.~\ref{eq:lfs}) passes
    through the supercooled, LFS-poor liquid.  Shown is the scenario of a
    lower critical point of the dynamical transition {at $\Tc$} passed by
    the unbiased dynamics at {a crossing temperature} $\Tx$ {with $\Tk\approx \Tx\approx\Tc$}.  Inset: Example for the determination
    of the coexisting LFS populations from the peak positions of histograms
    $p(n)$ evaluated at $\mu=\mu_\ast$ for $\Ts=0.60$ and
    $t_{\rm obs}=4.5\tau_\alpha$. {Finally, (c)(d)~and (e)~are sketches for alternative scenarios, with (f) (g) and (h) being their counterpart in the $T-\mu$ space: in (c,f) the coexistence region extends down to $T=0$;
    in (d,g)~ two critical points bound the coexistence region with the equilibrium line approaching the lower critical point in an ``avoided transition''  scenario; and in (e,h), the liquid undergoes a weakly first order transition at low temperature.} {At the onset temperature $T_{\rm o}$ and above, dynamical heterogeneities vanish and {in its vicinity} we schematically locate an upper critical point bounding the coexistence region.}}
  \label{fig:phase}
\end{figure*}

We draw together our results in a nonequilibrium phase diagram where both the equilibrium liquid and the nonequilibrium coexistence are represented \ft{[see Fig.\ref{fig:phase}(a), (b)]}.
To construct the \rev{nonequilibrium} phase diagram we determine the populations $\np$ and $\nr$ of
the coexisting LFS-poor and LFS-rich states, respectively. To this end we
produce histograms of the LFS population at $\mu=\mu_\ast$ and determine the
positions of the two peaks corresponding to each phase [see
Fig.~\ref{fig:phase}(b) inset]. These populations delimit the coexistence
region as shown in Fig.~\ref{fig:phase}(\ft{b}) for several sampling
temperatures. They agree within errors for different trajectory lengths and
system sizes ($N=216$ and $N=400$ particles). The unbiased equilibrium
dynamics with $\mu=0$ corresponds to a line in the $(T,n)$ plane as shown in red in
Fig.~\ref{fig:phase}(\ft{b}), any point away from this line corresponds to a
nonequilibrium state with $\mu\neq 0$. For the temperatures that we can
sample directly, the LFS population \rev{in the equilibrium liquid lies close to the spread of the data for the LFS-poor region}.


\section{Discussion}

\subsection{A lower critical point \rev{of the nonequilibrium phase transition}?}

The existence of a thermodynamic phase transition for glass forming
fluids is widely debated~\cite{cava09,chan10,bert11,stil88}. In the case of
the Kob-Andersen mixture, there are several pieces of numerical evidence
pointing towards a transition at temperatures substantially below the
Mode-Coupling transition temperature $T_\text{MC}\simeq0.435$:
(i)~Thermodynamic integration~\cite{scio99} provides, through an
extrapolation, an estimate for a finite Kauzmann temperature $\Tk$ at which
the configurational entropy vanishes. (ii)~As previously mentioned, relaxation
times can be thought to diverge in the vicinity of $\Tk$, providing a
dynamical description of the transition [see
Fig.~\ref{fig:basics}(a)]. (iii)~Recent equilibrium simulations~\cite{ozaw14} in the
presence of a variable concentration of pinned particles $c$ show that in the
limit of vanishing $c\to 0$ a transition between a more and a less ``glassy''
thermodynamic state (as indicated by the so-called overlap order parameter, see
\cite{ozaw14} and references therein) can be inferred at a nonzero temperature
close to $\Tk$. (iv)~Finally, the same overlap order parameter~\cite{bert2015} has
been used to show that static fluctuations exist for $T\lesssim 0.5$,
consistent with the existence of a random-\rev{field-like} critical point predicted by
effective field theories \cite{franz2013,biroli2014}. 

However, it is well known that the Vogel-Fulcher-Tamman fit is not the only --- nor even the best --- fit to the structural relaxation time \cite{heck08}. Others, 
which do not imply a thermodynamic transition,  
give as good a description \cite{elma09,tarj05}. Furthermore, the dynamical aspects, {in the sense of particle mobility} of the nonequilibrium transition that we study cannot be overstated, be {they} 
generated dynamically \cite{hedg09} or by structural averaging along trajectories as in our case. These nonequilibrium phase transitions lead to an inactive state whose time correlations do not decay on the simulation timescale. In other words, the inactive phase is a glass, consistent with the dynamic facilitation scenario {which posits that there is no thermodynamic phase transition.}

In the scenario we consider here, the thermodynamic behaviour of the system follows from the more
general description of its nonequilibrium phase diagram, where
the equilibrium limit corresponds to $\mu\to0$. We find \rev{compelling} numerical
evidence for a coexistence region between LFS-rich and LFS-poor trajectories,
at least in a the temperature interval $0.48\leqslant\Ts\leqslant0.74$, directly
probed by the biased simulations. The temperatures at which the distinction
between LFS-rich and LFS-poor trajectories ceases would identify an upper and
a lower critical point {to the nonequilibrium phase transition}, the locations of which are inferred by
extrapolation. In particular, the precise location of the lower critical point $\Tc$
with respect to the line \rev{corresponding to the equilibrium liquid} allows for a connection between the
putative thermodynamic transition at $\Tk$ and the dynamical phase transition
determined in the present work. In the following, we discuss our evidence for
a thermodynamic transition and also {consider} 
alternative scenarios.

\subsection{Numerical evidence for a thermodynamic transition and critical point}

To interpret our numerical results, {we first consider the extrapolation of the dynamical
chemical potential $\mu_\ast$ at coexistence}. In Fig.~\ref{fig:phase}(a) {we illustrate the phase diagram} in the $T-\mu$ plane, with the coexistence
region being {represented by} a line separating the LFS-poor from the LFS-rich
trajectories. Here we have plotted the different values of {chemical potential} $\mu_\ast$ {at coexistence} \rev{for different sampling temperatures $\Ts$,} scaled
with the trajectory length $\tobs$ and observe that they collapse onto a
common master curve. Down to the coldest sampling temperature $\Ts=0.48$,
$\mu_\ast(T)$ follows a linear behaviour, the extrapolation of which to
$\mu_\ast=0$ provides a second estimate for a 
crossing temperature
{$\Tx(\mu_\ast\to0)\simeq0.37$}, {23\% higher than the} estimate for the Kauzmann temperature $\Tk\simeq 0.3$. \ft{We observe [inset of Fig.~\ref{fig:phase}(a)] that the position of the maximum of the susceptibility $\chi(\mu)$ moves from very large values of $\mu_\ast$ at high temperature (i.e. very rare fluctuations) towards $\mu=0$ (i.e. typical equilibrium fluctuations) for lower temperatures. This is accompanied by a non-monotonic behaviour of the peak amplitude, suggesting critical-like fluctuations at some finite high temperature (that for physical reasons we relate to the onset of glassy dynamics) and a low temperature $\Tx$.} {If we only fit \ft{of} the coexistence points \ft{in the $T-\mu$ plane} inside the regime for which inherent states are well defined and the susceptibility peak $\chi$ is monotonically increasing as we decrease the temperature ($T\leq0.60$, see inset in Fig.\ref{fig:phase}(a)) the linear fit provides a lower estimate for $\Tx=0.33$ closer to the {estimate of the} Kauzmann {temperature}. }

 Our second piece of evidence {for a thermodynamic transition} comes from the population of LFS at
coexistence \rev{between the LFS-rich and poor phases}, $n_\ast(T)=\mean{n}_{T,\mu_\ast}$. {To do so we cast the phase diagram in the $T-n$ plane}. As shown in
Fig.~\ref{fig:phase}(b) (green {dashed} line), the dependence of $n_\ast$ {upon} 
temperature
is linear to a very good degree. Extrapolating towards lower temperatures, it
crosses the equilibrium ($\mu=0$) line \rev{at a ``crossing'' temperature $\Tx$} with
$n_\ast(\Tx)=\mean{n}_{\Tx,\mu=0}$, which implies $\np=\nr$. The crossing of
both lines occurs at {$\Tx(\np=\nr)\simeq 0.31$} close to the estimated
value for the Kauzmann temperature $\Tk\simeq0.30$.

{Our third piece of evidence stems from the inherent state energies.} We note that the inherent state energy of the LFS-rich configurations in
Fig.~\ref{fig:dist}(c) is approximately linear, 
\begin{equation}
\label{eq:phir}
\phir(T)=\phi_0+\gamma T,
\end{equation} with fitted $\phi_0\simeq-7.82$ (note that $\phi_0\approx\phik$) and
$\gamma\simeq0.18$ for the lowest sampling temperature. Assuming that the
vibrational free energy and \rev{that of the} inherent states still decouple, for any
transition the configurational temperatures (\emph{i.e.}, the slopes of the
configurational entropies of the LFS-rich and LFS-poor phases) have to agree. For a
putative \rev{continuous transition}
we equate the inherent state energy (\rev{using the equilibrium liquid Eq.~\ref{eq:mean} for the LFS-poor and Eq.~\ref{eq:phir} for the LFS-rich phase}) and find \rev{$\phi(\Tx)\simeq-7.76$}
with corresponding temperature \rev{$\Tx(\phir=\phip)\simeq0.33$}, which \rev{also provides an estimate for the temperature at which the two phases are indistinguishable, the nonequilibrium critical temperature $\Tc$.} 

{Finally, the information contained in the narrowing of the gap between the inherent state energies of the LFS-rich and LFS-poor phases can be translated into the $T-n$ phase diagram through additional numerical nonequilibrium simulations. The estimate of the inherent state energy of the LFS-rich phase at low temperatures is used (see Appendix~\ref{app:neq}) to select a representative population of configurations. We then use these as initial configurations for molecular dynamics simulations at temperature $T$ that explore a metastable basin and eventually relax to the equilibrium liquid state. We assume that the LFS-poor liquid nucleates within the LFS-rich liquid and therefore determine the characteristic time for nucleation $\tau_{\rm melting}$ and the LFS populations before and after nucleation. These populations are then employed as estimates at temperature for the coexisting LFS-rich and LFS-poor densities respectively (see Fig.\ref{fig:phase}(b)). We notice that our estimate for the gap $\nr-\np$ narrows from $\nr-\np=0.15(1)$ at $T=0.70$ to  $\nr-\np=0.09(1)$ at $T=0.40$. }

Although relying on extrapolations, these {four} pieces of numerical evidence
indicate that the dynamical transition from LFS-poor to LFS-rich {would} become
accessible for the equilibrium supercooled liquid provided that it is cooled
sufficiently slowly and crystallization does not interfere. From the numerical data we can bound the \rev{\textit{crossing}} temperature $\Tx$ at which \rev{the equilibrium (LFS-poor) liquid becomes undistinguishable from the low energy, LFS-rich amorphous phase, obtaining } 
${0.31}< \Tx<0.33$. \rev{This range is intriguingly close to previous estimates of the Kauzmann temperature $\Tk\simeq0.30$ at which a thermodynamic phase transition has been predicted \cite{scio99,ozaw14}}. \rev{Moreover, from the narrowing of the gap between the LFS-poor and LFS-rich coexisting phases we infer} that a lower critical point exists at a temperature $\Tc$, terminating the \rev{nonequilibrium} coexistence region between LFS-poor and LFS-rich
trajectories.

\subsection{The low temperature fate of the supercooled liquid}

Our numerical precision does not enable us to distinguish the equilibrium liquid and the LFS-poor phase, that is to say $\Tx\approx\Tc$. However, considering the precise location of \rev{the nonequilibrium critical point} $\Tc$ with respect to $\Tx$ we obtain three possible scenarios: First, the critical
temperature $\Tc>\Tx$ is higher, in which case no equilibrium transition
occurs but is passed closely (an avoided transition), {Fig. \ref{fig:phase}(d,g)}. At $\Tx$ the {equilibrium} system then crosses the line emanating from the {nonequilibrium} critical point where the fluctuations are maximal (sometimes called the ``Widom
line''~\cite{widom1965,xu05}), see Fig.~\ref{fig:phase}(d), analogously to some scenarios in softened kinetically constrained models \cite{elma13}. Second, in the case that
$\Tc=\Tx$ the {equilibrium} unbiased dynamics crosses through the critical point and a
continuous transition occurs as depicted in Fig.~\ref{fig:phase}(b). The final
possibility is that the {nonequilibrium} critical point lies \emph{below} the crossing temperature, 
$\Tc<\Tx$. In this case the equilibrium transition, {corresponding to the crossing of the coexistence region with a small but finite increase of the LFS concentration,} would be (weakly) first order as
sketched in Fig.~\ref{fig:phase}(e,h).

{Before concluding, we emphasize that our discussion concerns the linear extrapolation of the {dynamical} chemical potential $\mu_\ast$ and LFS population $n_\ast$ at coexistence, a supposition that the inherent state energy continues as indicated in Fig. \ref{fig:phi}, and that the relaxation time follows the VFT prediction. Other alternatives are possible, such as a more gradual approach to a state of very low configurational entropy, which could be met only at $T=0$. This could be related to ``defects'' in the amorphous order \cite{stil88}. Other scenarios have been suggested in which $\mu_\ast(T)$ approaches equilibrium ($\mu_\ast(T)=0$) more slowly at lower temperature, only reaching zero at $T=0$ \cite{jack10}, {as depicted in Fig.~\ref{fig:phase}(c,f)}. These would imply deviations in the relaxation time from the VFT behaviour. It is even possible that these effects lead to behaviour similar to that we discuss, but at a lower temperature than our extrapolations suggest.}

\section{Conclusions}

By means of numerical simulations employing trajectory sampling, we have
explored the connection between a dynamical phase transition and the low
temperature thermodynamics of a model 
atomistic glass former. We have 
{used} locally favoured structures 
as an order parameter to bias the simulations {in trajectory space} at a given sampling
temperature. {Thus, we have shown that this} is a powerful and reliable method to obtain equilibrium,
thermodynamic information (such as the inherent state energies and the
configurational entropy) down to {exceptionally} low temperatures. This is because
trajectory sampling facilitates the exploration of states that are
exponentially unlikely to be accessed in a conventional molecular dynamics
simulation.

Employing the probability distributions obtained, we show that the
dynamical phase transition in trajectory space implies a bimodal distribution
of inherent state energies: for the equilibrium dynamics most of the
configurations are compatible with a high energy, locally  favoured structure-poor state, the normal liquid. However, we
also reveal that an amorphous, LFS-rich, low energy state exists and is
accessible when the dynamical chemical potential $\mu$ is tuned away from
zero~\cite{spec12b}. 
We also determine the
temperature dependence of the energy gap between the two states, which
provides numerical evidence for a lower critical point at a non-zero temperature $\Tc$ where
the {high energy} (LFS-poor) phase would be indistinguishable from the {low energy} (LFS-rich) phase. This is
further corroborated by the temperature at which the first order dynamical
phase transition appears to cross the unbiased dynamics
($\mu=0$). Numerically, we find that the temperature $\Tc$ is close
to the 
Kauzmann temperature {estimated from our configurational entropy measurements}.
{We confirm our findings with particle swap methods that we have tailored for this system.}

Our results support the idea that the low temperature fate of supercooled
atomistic glass formers is determined by the fluctuations of the population of
local structures, {which couples with the reduction of configurational entropy}. There is now ample numerical {and experimental} evidence that particle
dynamics, at least for a class of 
 model
glassformers~\cite{cosl07,mali13fara,hock14,leoc12}, is correlated with the presence of
LFS. 
\rev{We observe a nonequilibrium phase transition between two phases in trajectory space, LFS-poor and LFS-rich. The former is, given our numerical precision, indistinguishable from the equilibrium liquid. The latter exhibits very slow dynamics \cite{spec12b}. Glassy phenomenology such as dynamical heterogeneities and the emergence of slow dynamics may be related to this nonequilibrium phase transition {in trajectory space. Our numerical data provides evidence that the transition may}
terminate at a lower critical point close to the Kauzmann temperature $\Tk$.}

The connections we identify between dynamical coexistence, the distinct basins in the energy landscape and the location of a low temperature nonequilibrium critical point close to the equilibrium ($\mu\to0$) path in the nonequilibrium phase diagram provide the ingredients for a novel understanding of the low temperature fate of supercooled liquids. In particular,
our work opens a potential avenue to unify the competing theories that
have been developed in order to understand the microscopic origin of slow
glassy dynamics: On one hand, the thermodynamic picture of a decreasing configurational entropy, vanishing at a finite temperature around $\Tk$, is compatible with the emergence of the low energy, low entropy, LFS-rich state that we have identified. On the other hand, 
{our results suggest that}
down to $\Tk$, this state {may be} 
in 
coexistence {in trajectory space} with the normal equilibrium liquid, and that the glassy phenomenology \rev{may be related to} the vicinity of the equilibrium line to 
{this} nonequilibrium phase coexistence. Within this approach, structural and dynamical aspects are combined, and play a complementary role. 
We emphasise that our conclusions rely on extrapolation. Other possibilities, including a {continuous decrease of} the configurational entropy such that it approaches zero only as temperature also approaches zero are possible. That is to say {our results do not exclude the possibility that}, there is no vanishing of configurational entropy or indeed any transition at $\Tk$ \cite{jack10}. Further alternatives include a suppression of any transition around $\Tk$ by ``defects'' \cite{stil88}. {Finally, despite the Kob-Andersen mixture being a good glass-former, it is prone at low temperatures $T<0.40$ to crystallisation, with a micro-phase separation that promotes the formation of an fcc crystal of large particles coexisting with the fluid {\cite{toxv2009}}. This is in our calculations anti-correlated with the growth of the LFS-population so that it identifies a different regime, but the mechanisms that lead to crystallisation may play a role in the meta-stability of the LFS-rich phase. }

Applying the method we have developed 
to other model glass formers with different fragilities and different structural signatures~\cite{hock14}, and comparing it with existing theories for critical behaviour in glasses~\cite{franz2013,nandi2014} will be the challenging subject of future investigations. 
{Particle-resolved experiments with colloids and granular materials \cite{ivle} provide the natural means by which our predictions may be verified experimentally, as these experiments provide the same data as to the simulations we have used here. Progress has been made in this direction, with 
identification of the dynamical phase transition in trajectory space to an LFS-rich phase \cite{pinc16}. It is thus possible to directly investigate the phenomena we have identified here.

More generally, provided an LFS can be identified with a given system, we expect it to be possible {to} observe the kind of behaviour we see here. In particular we expect that other Lennard-Jones and hard sphere systems should exhibit similar behaviour, as LFS are known for these systems \cite{roya15}. In metallic glasses, LFS have also been found \cite{chen11,hira13,liu13}.  Other glassformers in which local structure has been investigated include oxides and other inorganic materials \cite{salm13}. {A challenging direction of research would be to apply trajectory sampling to simple models of silica (such as the so-called BKS model \cite{vanbeest90} or the NTW model \cite{cosl09}), where a crossover from fragile to strong behaviour is observed \cite{saika01}, in order to identify its relationship with the dynamical phase transition discussed in the present work}. {Beyond that, it is important to observe that} organic molecules and even polymers at the present time form an open challenge as LFS have yet to be identified. {For some coarse-grained models (such as the Lewis-Wahnstr\"om model for orto-terphenyl \cite{wahn1993}) it would be possible to identify local structure and its relation to the dynamics following an analysis similar the present work.}
 {However, the greatly increased number of degrees of freedom of {molecular} systems means that convincingly determining an LFS still constitutes a considerable task, and other forms of amorphous order may prove more amenable to identify the kind of phenomena we present here  \cite{hock12,camm12epl,cubu15,roya15physrep}. We stress however that we expect the behaviour we see to be generic to a wide range of materials.}


\acknowledgments

The authors are grateful to Ludovic Berthier, Juan P. Garrahan, Robert L. Jack, Walter Kob and
John Russo for helpful comments. Stephen Williams is gratefully acknowledged
for helpful discussions and preliminary results. CPR acknowledges the Royal
Society for funding and Kyoto University SPIRITS fund. FT and CPR acknowledge
the European Research Council (ERC consolidator grant NANOPRS, project number
617266). This work was carried out using the computational facilities of the
Advanced Computing Research Centre, University of Bristol.

\appendix
\section{Ghost-particle Monte-Carlo }
\label{app:gMC}
{We compare the inherent state energies obtained from trajectory sampling to equilibrium values obtained \textit{via} a modified Monte-Carlo technique inspired by the ghost particle insertion method \cite{ashton2011} and the employment of swap Monte Carlo techniques in order to equilibrate supercooled liquids closer to the glass transition.}

{The Kob-Andersen binary mixture is characterised by high density and relatively hard interactions which make traditional particle-swap techniques highly inefficient. Particle-swaps have appeared to be much more effective for ternary mixtures \cite{guti15}, soft interactions \cite{grigera2001}, or very polydisperse hard-spheres \cite{bert16}.}

{Here we choose to improve the sampling efficiency in the Kob-Andersen mixture introducing a third species in our system. At any time step, the system is split into $N-1$ particles interacting via the standard non-additive Kob-Andersen interactions (see Sec.~\ref{sec:model}) and a so-called ``ghost'' particle $g$. This can be both a large particle A or a small particle B and it interacts with any other particle $i$ via a capped potential:}
\begin{equation}
V_{g,i}^{s}	=\min(	V_{\gamma_g, \gamma_i}(r_{gi}), V_{\rm max}^{s})
\end{equation}
{where $\gamma_g$ and $\gamma_i$ correspond to the $A$ or $B$ types of particle $g$ and $i$ while $s$ is a discrete variable associated to the state of particle $g$. We define a discrete space of $N_{\rm stages}$ states corresponding to different values of the maximum allowed energy $V_{\rm max}^{s}$. During the simulation, we perform a Markov-chain in the space of states, with transitions between state $s$ and $s'=s\pm 1$ governed by a Metropolis acceptance probability:}
\begin{equation}
\label{eq:stagemove}
	p_\text{acc}=\min \left\{1,\exp\left(-\frac{1}{T} \sum_{i=1}^{N}V_{g,i}^{s'}-V_{g,i}^{s}\right)\right\}
\end{equation}
{We then implement the following algorithm:
\begin{enumerate}
	\item We perform $N$ attempts of local Monte-Carlo moves following the usual Metropolis prescription.
	\item We perform a Monte-Carlo step according to Eq.~\ref{eq:stagemove} and modify the discrete state variable $s$ and the corresponding potential cap $ V_{\rm max}^{s}$ associated to the ghost particle $g$.
	\item When the discrete variable $s$ assumes value $N_{stages}$, the ghost particle becomes a normal particle. We therefore store the corresponding configuration, and the ghost status $s=0$ is assigned to a new, randomly selected particle.
	\item We repeat $1-3$ in order to obtain equilibrated configurations, whose inherent state energies are evaluated using the FIRE algorithm (see Sec.~\ref{sec:model} above), while trajectory sampled configurations appear to sample energies that are significantly lower.
\end{enumerate}}

{Our choice for the discrete set of states is the following:
\begin{center}
  \begin{tabular}{r|C{1cm} C{1cm} C{1cm} C{1cm} C{1cm}}
    s & 0 & 1 & 2 & 3 &4 \\
    \hline
    $ V_{\rm max}^{s}/\epsilon_{AA}$ & 3 &4 &5 &12& $+\infty$
  \end{tabular}
\end{center}
The idea behind the algorithm is to facilitate the particle motion through lowering the energy barriers for cage breaks locally while preserving detailed balance. As shown in Fig.~\ref{fig:phi}, the technique produces reliable estimates of the inherent state energies down to $T=0.40$. At $T=0.375$ the system still appears to be trapped in a glassy state, within the considered computation time (256 CPU hours, $4\times10^7$ Monte Carlo sweeps) and trajectory sampling  provides significantly lower inherent state energies.}

{
\section{Nonequilibrium relaxation from LFS-rich configurations}
\label{app:neq}
In order to provide an estimate for the coexisting LFS-poor and LFS-rich populations at temperatures lower than lowest sampling temperature $T_s=0.48$ employed in the trajectory sampling technique, we perform conventional molecular dynamics simulations starting from a selected sample of initial configurations.

We first determine the expected average inherent state energy for the LFS-rich phase $\phir(T)$ at temperature $T$ via reweighting: see section \ref{sec:reweight} and Fig.~\ref{fig:phi}. We then extract from our population of sampled trajectories a number $L=70$ central configurations of $N=216$ particles whose energy is within the interval $[\phir-\sigma, \phir+\sigma]$ where $\sigma=0.006$ is chosen as the typical width at half-height of the LFS-rich peak in the reweighted probability distribution $p(\phi, \mu=\mu\ast)$ (see inset of Fig.\ref{fig:dist}(c)). We measure the relaxation from the the LFS-rich phase to the LFS-phase with a constant temperature protocol, where the system is coupled to a Nose-Hoover thermostat at temperature $T$ whose characteristic damping time is set to a tenth of the structural relaxation time.
}
{Following this protocol, we produce trajectories of duration $4\tau_\alpha$ temperatures $T=0.44,0.42,0.40$ and measure the population of LFS $n$ along the trajectories. For every trajectory we model the time evolution of the LFS population as a two state function $n(t)=A+B\tanh{(t-\tau_{\rm melting})/C}$ and fit the model to the data to obtain per-trajectory estimates of the time $\tau_{\rm melting}$ needed to melt the LFS-rich state. For every trajectory we then estimate the LFS-rich coexisting population with the time average of the LFS-population in the interval $[0,\tau_{\rm melting}]$ and also the LFS-poor population from the remaining part of the nonequilibrium trajetcory.

Averaging over all the initial conditions, we obtain mean and standard errors as represented in Fig.~\ref{fig:phase}(b) in pink (NVT protocol) symbols. The estimates for the LFS-poor phase (despite not resulting from equilibrium runs) are at most a 3\% higher than the prediction for the equilibrium liquid, Eq.\ref{eq:lfs} .
}

%

\end{document}